\documentclass[11pt,preprint]{aastex}
\usepackage{color}
\def\lsim{\raise0.3ex\hbox{$<$}\kern-0.75em{\lower0.65ex\hbox{$\sim$}}}
\def\gsim{\raise0.3ex\hbox{$>$}\kern-0.75em{\lower0.65ex\hbox{$\sim$}}}

\begin{document}
\title{Cluster Formation Triggered by Filament Collisions in
Serpens South
}
\author{
Fumitaka Nakamura\altaffilmark{1,2,3,4},
Koji Sugitani\altaffilmark{5}, 
Tomohiro Tanaka\altaffilmark{6},
Hiroyuki Nishitani\altaffilmark{2}, 
Kazuhito Dobashi\altaffilmark{7}, 
Tomomi Shimoikura\altaffilmark{7}, 
Yoshito Shimajiri\altaffilmark{8}, 
Ryohei Kawabe\altaffilmark{1} 
Yoshinori Yonekura\altaffilmark{9}, 
Izumi Mizuno\altaffilmark{2,11}, 
Kimihiko Kimura\altaffilmark{6},
Kazuki Tokuda\altaffilmark{6},
Minato Kozu\altaffilmark{6},
Nozomi Okada\altaffilmark{6},
Yutaka Hasegawa\altaffilmark{6},
Hideo Ogawa\altaffilmark{6},
Seiji Kameno\altaffilmark{12}, 
Hiroko Shinnaga\altaffilmark{1}, 
Munetake Momose\altaffilmark{13},
Taku Nakajima\altaffilmark{14},
Toshikazu Onishi\altaffilmark{6},
Hiroyuki Maezawa\altaffilmark{6},
Tomoya Hirota\altaffilmark{1},
Shuro Takano\altaffilmark{2,3}, 
Daisuke Iono\altaffilmark{1}, 
Nario Kuno\altaffilmark{2,3,15}, 
Satoshi Yamamoto\altaffilmark{16}
}
\altaffiltext{1}{National Astronomical Observatory, Mitaka, Tokyo
181-8588, Japan; fumitaka.nakamura@nao.ac.jp}
\altaffiltext{2}{Nobeyama Radio Observatory, Minamimaki, Minamisaku, 
Nagano 384-1305, Japan}
\altaffiltext{3}{The Graduate University for Advanced Studies
(SOKENDAI), 2-21-1 Osawa, Mitaka, Tokyo 181-0015, Japan}
\altaffiltext{4}{
Kavli Institute for Theoretical Physics, University of California, Santa
Barbara, CA 93106-4030}
\altaffiltext{5}{Graduate School of Natural Sciences, 
Nagoya City University, Mizuho-ku, Nagoya 467-8501, Japan}
\altaffiltext{6}{Department of Physical Science, Osaka Prefecture
University, Gakuen 1-1, Sakai, Osaka 599-8531, Japan}
\altaffiltext{7}{Department of Astronomy and Earth Sciences, 
Tokyo Gakugei University, Koganei, Tokyo 184-8501, Japan}
\altaffiltext{8}{Laboratoire AIM, CEA/DSM-CNRS-Universit$\acute{e}$
Paris Diderot, IRFU/Service d'Astrophysique, CEA Saclay, F-91191 Gif-sur-Yvette, France}
\altaffiltext{9}{Center for Astronomy, Ibaraki University, 2-1-1 Bunkyo, 
Mito, Ibaraki 310-8512, Japan}
\altaffiltext{10}{Institute of Astrophysics and Planetary Sciences, 
Ibaraki University, Bunkyo 2-1-1, Mito 310-8512, Japan}
\altaffiltext{11}{Department of Physics, Faculty of Science, 
Kagoshima University, 1-21-35 Korimoto, Kagoshima, Kagoshima 890-0065, Japan}
\altaffiltext{12}{Joint ALMA Observatory, Alonso de Crdova 3107 Vitacura, Santiago, Chile}
\altaffiltext{13}{Institute of Astrophysics and Planetary Sciences, 
Ibaraki University, Bunkyo 2-1-1, Mito 310-8512, Japan}
\altaffiltext{14}{Solar-Terrestrial Environment Laboratory, Nagoya University, Furo-cho,
Chikusa-ku, Nagoya, Aichi 464-8601, Japan}
\altaffiltext{15}{Department of Physics, Graduate School of Pure and Applied Sciences, 
The University of Tsukuba, 1-1-1 Tennodai, Tsukuba Ibaraki 305-8577, Japan}
\altaffiltext{16}{Department of Physics, Graduate School of Science, 
The University of Tokyo, Tokyo 113-0033, Japan}
\begin{abstract}
The Serpens South infrared dark cloud consists of several filamentary 
ridges, some of which fragment into dense clumps. On the basis of 
CCS ($J_N=4_3-3_2$), HC$_3$N ($J=5-4$),  
N$_2$H$^+$ ($J=1-0$), and SiO ($J=2-1, v=0$) observations, we investigated the 
kinematics and chemical evolution of these filamentary ridges. We find that 
CCS is extremely abundant along the main filament in the
 protocluster clump.
We emphasize that Serpens South is the first {\it cluster-forming}
 region where extremely-strong CCS emission is detected.
The CCS-to-N$_2$H$^+$ abundance ratio is estimated to be
 about 0.5 toward the protocluster clump, whereas it is about 3 in the
other parts of the main filament.
We identify six dense ridges with different $V_{\rm LSR}$.
These ridges appear to converge toward the protocluster clump, 
suggesting that the collisions of these ridges may
have triggered cluster formation.
The collisions presumably happened within a few $\times \ 10^5$ yr
because CCS is abundant only in such a short time. 
The short lifetime agrees with the fact that the number fraction of Class I objects, whose typical
 lifetime is $0.4 \times \ 10^5$ yr, is
 extremely high as about 70 percent in the protocluster clump.  
In the northern part, two ridges appear to have 
 partially collided, forming a V-shape clump. 
In addition, we detected 
strong bipolar SiO emission that is due to the
 molecular outflow blowing out of the protostellar clump, as well as extended weak
 SiO emission that may originate from the filament collisions.
\end{abstract}

\section{Introduction}
\label{sec:intro}

Over the last few decades, considerable progress has been made in the understanding 
of star cluster formation \citep[e.g.,][]{lada03}.
For example, radio and infrared observations have revealed 
that star clusters form in parsec-scale dense molecular clumps 
with masses of order of $10^2-10^3 M_\odot$
\citep{lada03,ridge03,battersby10}.
In the cluster-forming clumps, stellar feedback plays a crucial 
role in the structure formation of the clumps \citep{sandell01}.
The clumps tend to keep quasi-virial equilibrium 
states at least for several  free-fall times, by
sustaining supersonic turbulence supplied by 
stellar feedback \citep[e.g.,][]{tan06,li06,matzner07,nakamura14}.

In contrast, the initial conditions of cluster formation remain to be elucidated
because of the lack of the observational characterization. 
This is because once active star formation is initiated,  
stellar feedback such as protostellar outflows and radiation
rapidly shapes their surroundings, making it 
difficult to track back the physical conditions of the clumps prior to
active cluster formation, i.e., pre-protocluster clumps 
(or,  precluster clumps, in short). 
According to recent numerical simulations, 
the initial conditions such as density profile, magnetic 
field distribution, and turbulent Mach number determine
some properties of forming clusters 
\citep{girichidis12,li10} 
because these factors control the fragmentation process.
Therefore, it is important to search for molecular clouds where 
active cluster formation have just been initiated or is just about to 
be initiated.

Recently, \citet{tanaka13} 
found that the Serpens South infrared dark cloud (IRDC, $d\sim 415$ pc) is the region 
that contains both extremely-young protocluster and precluster clumps.
The Serpens South IRDC is a filamentary molecular cloud that 
contains several filamentary ridges 
 \citep{gutermuth08,andre10,maury11,kirk13,arzoumanian13}.
The central clump is associated with a very young 
($\lesssim $ 0.5 Myr) cluster of low-mass protostars \citep{gutermuth08} and 
is most massive with about 230 $M_\odot$ \citep{tanaka13}.
A northern clump shows no sign of active star formation, although its
mass ($\sim 200 \ M_\odot$) and size are comparable to those of the central protocluster clump, 
i.e., this northern clump is the precluster clump.
Both protocluster and precluster clumps have  virial ratios as small as 0.2, 
indicating that the internal turbulent motions cannot support the whole clumps against gravity.
In spite of the small virial ratios, the clumps do not exhibit 
rapid infall motions, and therefore their global contraction
is likely to be delayed by the moderately-strong magnetic fields that are 
 perpendicular to the filament axis. 
From the Chandrasekhar-Fermi method,
 the field strength is estimated to be a few $\times$ 
$10^2 \ \mu$G, close to the critical value above which 
the cloud can be supported against collapse
\citep{sugitani11}.

In the present paper, we aim to unveil how cluster formation has been
initiated in Serpens South, by investigating the kinematics of filamentary 
ridges on the basis of CCS and N$_2$H$^+$ observations.
CCS and N$_2$H$^+$ are known to be abundant in early and late phases of
prestellar evolution, respectively, and therefore are appropriate to inspect 
the evolutionary stages of young star-forming regions
\citep[e.g.,][]{hirota09}.

\section{Observations and Data}
\label{sec:obs}

We carried out CCS $J_N=4_3-3_2$ (45.379033 GHz) and HC$_3$N $J=5-4$ (45.490316 GHz) 
mapping observations toward Serpens South in 2013 May with the Nobeyama 45-m telescope. 
We use
a new 45 GHz band, dual-polarization receiver, Z45 \citep{tokuda13}.
The main beam efficiency and Half Power Beam Width (HPBW) were 
$\eta_{\rm Z45} \simeq $ 0.7 
and $\Theta \simeq 37''$, respectively.
At the back end, we used the 4 sets of 4096 channel 
SAM45 spectrometer whose frequency resolution was set to 3.81 kHz
($\approx 0.025 $ km s$^{-1}$ at 45 GHz), 
The intensity calibration was made by observing the cyanopolyyne peak of
TMC-1 using the S40 receiver + AOS system.

We also carried out SiO $J=2-1$ (86.847010 GHz) observations in 2013
May using the TZ + SAM45 receiver system of the 45-m
telescope. 
The main beam efficiency and HPBW were 
$\eta_{\rm TZ} \simeq $ 0.36 and 19$''$ at 86 GHz, respectively
\citep{nakajima13}.

The typical system temperatures during the observations were 150 K and
120 K for Z45 and TZ, respectively.
The telescope pointing was checked every 1 hour by observing a 
SiO maser source, IRC+00363, and the typical pointing offset was
better than 3$''$ during the whole observing period.
By applying a convolution scheme with a spheroidal function,  
the final maps were obtained by combining x-scan and y-scan OTF data 
into single maps with grid sizes of 15$''$ and 7.5$''$ for the Z45 and TZ
observations, respectively.
The resultant effective resolutions of the maps were about 49$''$ and
25$''$ for Z45 and TZ, respectively.
In the following, the intensities are shown
in the main beam brightness temperature units, 
$T_{\rm mb} = T_{\rm A}^*/\eta_{\rm Z45}$ or $T_{\rm A}^*/\eta_{\rm TZ}$ , 
where $T_{\rm A}^*$ is the antenna temperature corrected 
for the atmospheric attenuation.
The rms noise levels are 0.32 K and 0.24 K at the velocity resolution
of 0.05 km s$^{-1}$ for Z45 and TZ, respectively.

For the N$_2$H$^+$ data, we used the data presented in \citet{tanaka13}.

\section{Global Gas Distribution and Kinematics}
\label{sec:results}

\begin{figure}[h]
\epsscale{1.0}
\plotone{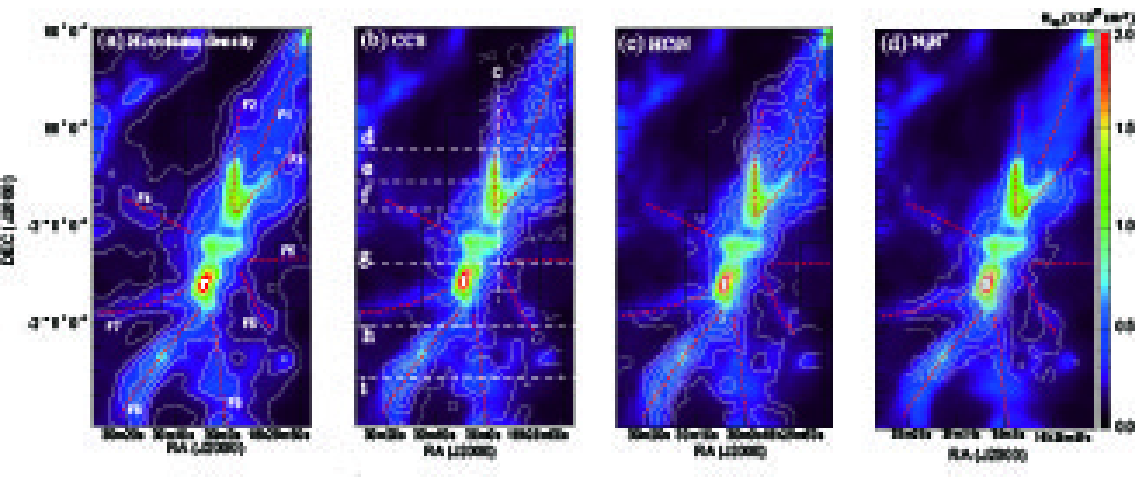}
\caption{(a)
The H$_2$ column density map (see Tanaka et al. 2013 for detail.)
The positions of nine ridges are indicated with the dashed curves.
The contours start at $0.12 \times 10^{23}$ cm$^{-2}$ with a logarithmic
 interval of $\log [\Delta N_{\rm H_2} / (1.2\times 10^{23} \ {\rm cm}^{-2})] =
 $ 0.2.
(b) CCS ($J_N=4_3-3_2$) integrated intensity contour map 
in the velocity range from $V_{\rm LSR}$ = 5.5 to 9.0 km s$^{-1}$.
The image is overlaid on the H$_2$ column  density map.
The contour level starts from 0.33 K km s$^{-1}$ with an interval of
 0.33 K km s$^{-1}$, corresponding to the 3 $\sigma$ noise level.
(c) HC$_3$N ($J=5-4$) integrated intensity contour map 
in the velocity range from $V_{\rm LSR}$ = 5.5 to 9.0 km s$^{-1}$.
The image is overlaid on the H$_2$ column  density map.
The contour level starts from 0.3 K km s$^{-1}$ with an interval of
 0.4 K km s$^{-1}$.
(d) N$_2$H$^+$ ($J=1-0$) integrated intensity contour map overlaid on 
the H$_2$ column density map.
The contour level starts from 2 K km s$^{-1}$ with an interval of 2 K km s$^{-1}$.
}  
\label{fig:herschel}
\end{figure}

In Serpens South, several less-dense filamentary ridges appear to
converge toward the protocluster clump \citep[see also][]{myers09}. 
In Figure \ref{fig:herschel}(a), we show the positions of nine representative 
filamentary ridges that are overlaid on 
the H$_2$ column density map obtained by
{\it Herschel} data with 36$''$.3 effective resolution \citep[see][]{tanaka13}.
The positions of the ridges are determined by eye from the {\it
Herschel} map.
The identified ridges are named as F1, F2, $\cdots$, F9 in Figure \ref{fig:herschel}(a).

F2 and F8 trace the main filamentary cloud.
F2 and F3 appear to be connected with each other at their southern tips, 
creating the V-shape precluster clump.
F1 is located between F2 and F3.
F2, F3, and F8 are essentially the same as those identified by \citet{kirk13} 
who found the large velocity gradient of $2.2 \pm 0.3$ km s$^{-1}$ 
pc$^{-1}$ along F8, where we adopted the distance
of 415 pc.
F7 is extended toward the eastward direction and 
can also be recognized in the {\it Spitzer} IRAC image 
\citep[see Figure 1 of ][]{gutermuth08}.

The CCS, HC$_3$N, and N$_2$H$^+$ integrated intensity maps show
that 
their emission traces reasonably well the dense parts of the cloud
(Figures. \ref{fig:herschel}(b) though \ref{fig:herschel}(d)).
F1, F2, F3, F7, F8, and F9 are detected in N$_2$H$^+$, HC$_3$N, and/or
CCS, 
and thus contain dense gas with densities of $10^4-10^5$ cm$^{-3}$.
Other ridges, F4, F5, and F6, are not clearly detected in CCS, HC$_3$N,
and N$_2$H$^+$.
The CCS emission tends to be strong along the main filamentary cloud
F2$+$F8. 
In particular, the CCS emission is extremely strong along F2, 
whereas it is weak in the protocluster clump
where the N$_2$H$^+$ emission is strongest.
The CCS column density is estimated to be about 
$\sim 1\times 10^{14}$ cm$^{-2}$ toward the precluster clump. Such an
extremely
high column density of CCS is typical of the Carbon-Chain Producing
Regions suggested by \citet{hirota09}.

The CCS fractional abundances  are estimated to be about 
$1\times 10^{-9}$, $1\times 10^{-10}$,  and $1\times 10^{-9}$ toward the
dense part of F8,
protocluster clump, and precluster clump, respectively, 
where we adopted the excitation temperature of 5 K
and LTE approximation \citep[e.g.,][]{wolkovitch97,lai00}. 
We note that a change in excitation temperature of 1K
causes a change in column density and abundance of 30 \%.
The CCS fractional abundance in the protocluster
clump is of the order of magnitude smaller than the other parts of the
main filament.
The [CCS]/[N$_2$H$^+$] abundance ratio are also estimated to be
3.0, 0.4, and 2.5, for F8, protocluster clump, and precluster clump,
respectively, where N$_2$H$^+$ abundances are adopted from
\citet{tanaka13}.
In other words, the distribution of CCS tends to be anti-correlated 
with that of N$_2$H$^+$.
According to the chemical evolution
\citep[e.g.,][]{suzuki92,marka12},
CCS is abundant only in the early phase of prestellar evolution 
($\lesssim $ a few $\times $ $10^{5}$ yr), whereas
N$_2$H$^+$ becomes abundant in the late phase.
Thus, CCS and N$_2$H$^+$ can often been used as an age indicator of
dense cores \citep[e.g.,][]{degregorio06,sakai07,hirota09,devine11}.
The CCS-rich F2 and F8 ridges are likely to be chemically 
and dynamically young.
HC$_3$N distribution is roughly similar to that of CCS.

To see the global kinematic motions of the filamentary ridges, 
we show in Figures \ref{fig:vel} (a) and \ref{fig:vel} (b) the centroid
velocity maps 
of N$_2$H$^+$ and CCS, respectively, which are overlaid on the {\it
Herschel} 
H$_2$ column density contour map.
The CCS and N$_2$H$^+$ centroid velocity maps are in good agreement with
each other, except at the central clump.
The maps show that the main filamentary ridge, F8, is blueshifted
from the systemic velocity of 
the central protocluster clump ($V_{\rm sys} \simeq $ 7 km s$^{-1}$).

On the other hand, other ridges, F2, F3, F7, and F9, are 
redshifted from $V_{\rm sys}$ (Figures. \ref{fig:vel} (a) and
\ref{fig:vel} (b).)
The velocity structures in the northern part are somewhat complicated.
The typical difference in $V_{\rm LSR}$ is around 0.5 $-$
0.7 km s$^{-1}$ in the northern part.
The dynamical time of the filament 
suggests that the filaments are formed with short spacing of 0.4 $-$ 0.5
pc ($\approx 0.5 \ {\rm Myr} \times (0.5 - 0.7)\ {\rm km \ s^{-1}} / \cos
45^\circ$), 
where we assume that the collision velocity is equal to the velocity
difference and this region has an inclination of 45$^\circ$.
Here we adopt the dynamical time of about $ 0.5$ Myr, which is the
lifetime of Serpens South protocluster.  
This suggests that the filaments do not form by gravitational fragmentation,
instead by external events such as turbulent compression, and the
densities become as high as 10$^4$ cm$^{-3}$ within a few $\times$
$10^5$ Myr.

In fact, the CCS line profiles shown in Figures
\ref{fig:vel} (c) through \ref{fig:vel} (f) consist of a couple of components 
with different velocities in several positions around the V-shape
clump. 
The CCS line width of each component is typically 0.6 $-$ 0.8 km s$^{-1}$, which
is about 1.5 $-$ 2 times larger than the typical CCS line widths 
(0.3 $-$ 0.6 km s$^{-1}$) of more quiescent clouds such as Taurus \citep{suzuki92}.



The main ridge F8 has a significant velocity gradient that can be 
interpreted as an infall motion toward the protocluster 
clump \citep{kirk13}, although the northern part of the filament
does not show similar velocity gradients.
On the other hand, CCS is weak along F7 and F9.
It is worth noting that Serpens South is a really unique 
{\it cluster-forming} region in terms of the extremely-high fractional 
abundance of CCS and other carbon-chain molecules such as HC$_7$N 
\citep{friesen13}.
The CCS line widths are also wider than those of nearby CCS-rich dark
clouds such as Taurus.

\section{A Scenario of Triggered Cluster Formation in  Serpens South}
\label{sec:dis2}

From these characteristics, we propose the following scenario of 
cluster formation in Serpens South, as in Figure \ref{fig:scenario}.

In Serpens South, the magnetic field is relatively strong and plays an
important role in cloud dynamics \citep{sugitani11}. In the parent cloud, 
multiple filamentary ridges were created due to internal MHD turbulence
and/or large-scale flows.
These filamentary ridges were preferentially perpendicular to 
the global magnetic field lines and 
moved along the magnetic field lines.
The filaments also fragmented into several subclumps. Since the filaments
were likely to be magnetically supported \citep{sugitani11}, 
the fragmentation was presumably triggered by ambipolar diffusion 
accelerated in a turbulent environment \citep{nakamura08,tanaka13}.
The collisions tend to make the effective magnetic support
reduced and inject additional turbulence in the dense parts.
Consequently, cluster formation was initiated in the central clump.

It is difficult to identify which ridges are paired with which
because the velocity structure is complicated particularly in the northern part.
But, we suggest that F2 and F8 were from the same filament 
because CCS is abundant along these ridges.
This identification agrees with the structure identified by \citet{kirk13}.
\citet{kirk13} also suggested that F3 is paired with F9. If this is the case, 
F1 and F7 might be from the same filament. 
But, there is a possibility that F1 and F3 may be paired with F9 and F7, respectively.

In any of these cases, at least three filamentary ridges
may have collided at the position of the protocluster clump.
The collision may have happened within a few
$\times$ $10^5$ yr because most of the protostars in 
the dense clump are the Class I objects whose typical lifetime
is around 0.4 Myr \citep{evans09}.  This timescale  
is in good agreement with that inferred from the chemical 
evolution of CCS.
Two subclumps formed along the different ridges F2 and F3
have just collided with each other to form the V-shape clump, 
which may be on the verge of future cluster formation.
Such dynamical event may have caused the CCS line widths to be 
wider than those in more quiescent dark clouds.

\begin{figure}[h]
\epsscale{1}
\plotone{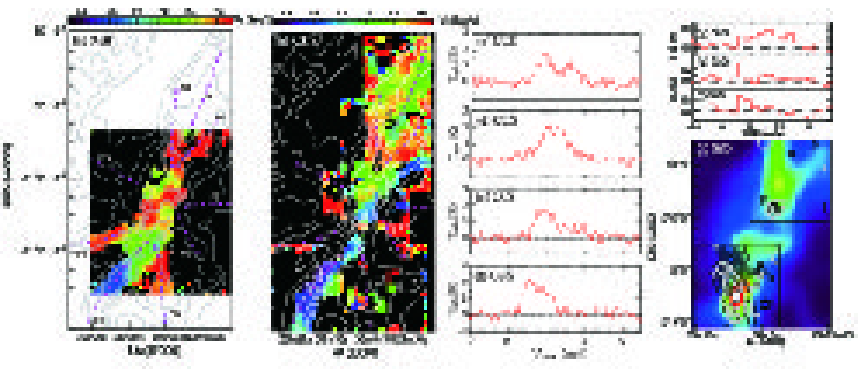}
\caption{
(a) Centroid velocity map for N$_2$H$^+$ ($J=1-0$) across Serpens South.
The N$_2$H$^+$ centroid velocity map was obtained 
by fitting the seven hyperfine lines with Gaussian profiles \citep{tanaka13}.
(b) same as panel (a) but for CCS ($J_N=4_3-3_2$).
For panels (a) and (b), the contours are the same as those in
 Fig. \ref{fig:herschel} (a).
(c) through (f) CCS line profiles at four positions that are designated 
in Fig. \ref{fig:vel}b with circles.
(g) SiO ($J=2-1$, $v=0$) line profile at the position g in panel (j).
(h) Same as panel (g) but for the position h.
(i) Same as panel (g) but for the position i.
(j) SiO ($J=2-1$) integrated intensity contour map
overlaid on  the H$_2$ column  density map.
The black and gray contours show the blueshifted and redshifted
 components integrated over $-10 \sim 7$ km s$^{-1}$ and 
$8 \sim 20$ km s$^{-1}$, respectively. 
The contour level starts from 1 K km s$^{-1}$ with an interval of
1 K km s$^{-1}$, corresponding to the 3.5 $\sigma$ and 2.5 $\sigma$
 noise levels for the blueshifted and redshifted components,
 respectively.
The color image is the same as that shown in Figure \ref{fig:herschel}.
}  
\label{fig:vel}
\end{figure}

\begin{figure}[h]
\plotone{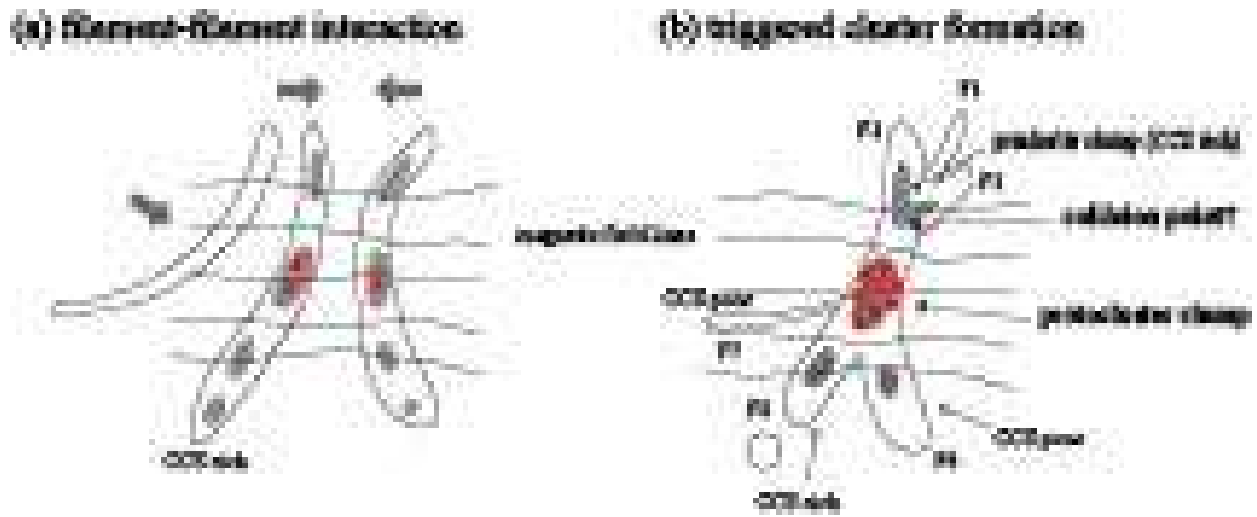}
\caption{Possible scenario of cluster formation triggerd by filament
 collisions in Serpens South. 
(a) at least three filamentary flows are on the verge of the 
collisions along the global  magnetic field lines.
The filaments fragment into several subclumps due to
 turbulence-accelerated ambipolar diffusion. 
Several stars may have formed in the subclumps before the collisions, and
these stars are now observed as Class II sources.
(b) They collided and crossed at the center where cluster formation was
triggered.  For the northern clump, the collision has just 
happened at its southern tip.
}
\label{fig:scenario}
\end{figure}

To search for further evidence of the filament collisions, we carried out
SiO ($J=2-1$, $v=0$) observations.
The SiO emission is sometimes used as a shock tracer of 
the cloud-cloud collision \citep[e.g.,][]{jimenez-serra10}.
Figure \ref{fig:vel} (j) shows the SiO integrated intensity map. 
We detected emission from both the central protocluster and northern
precluster clumps.
We note that \citet{kirk13} reported no detection of SiO emission on the
basis of Mopra 22-m observations. This apparent discrepancy comes from
the fact that the sensitivity of our map is much better than that of \citet{kirk13}.
The emission associated with the central clump is likely to
be due to a protostellar outflow, because 
the emission shows a well-defined  bipolar feature.
In addition, we detected weak emission lines at several positions
as presented in Figures \ref{fig:vel} (g) through \ref{fig:vel} (d).
These weak SiO lines might contain contribution of compressional heating by
the ongoing filament collision at around $V_{\rm LSR} \sim 7$ km s$^{-1}$.
However, the higher velocity component ($V_{\rm LSR} \gtrsim 10$ km
s$^{-1}$) near the southern tip of the northern clump presumably
originates from the interaction between the northern clump and the redshifted CO lobe 
identified by \citet{nakamura11} as R2, because the SiO emission
overlaps well with the northern part of the redshifted CO lobe whose
velocity range is $V_{\rm LSR} \sim 10-30$ km s$^{-1}$.
Further justification is needed to confirm the origin of the weak SiO emission.

\begin{figure}[h]
\epsscale{0.7}
\plotone{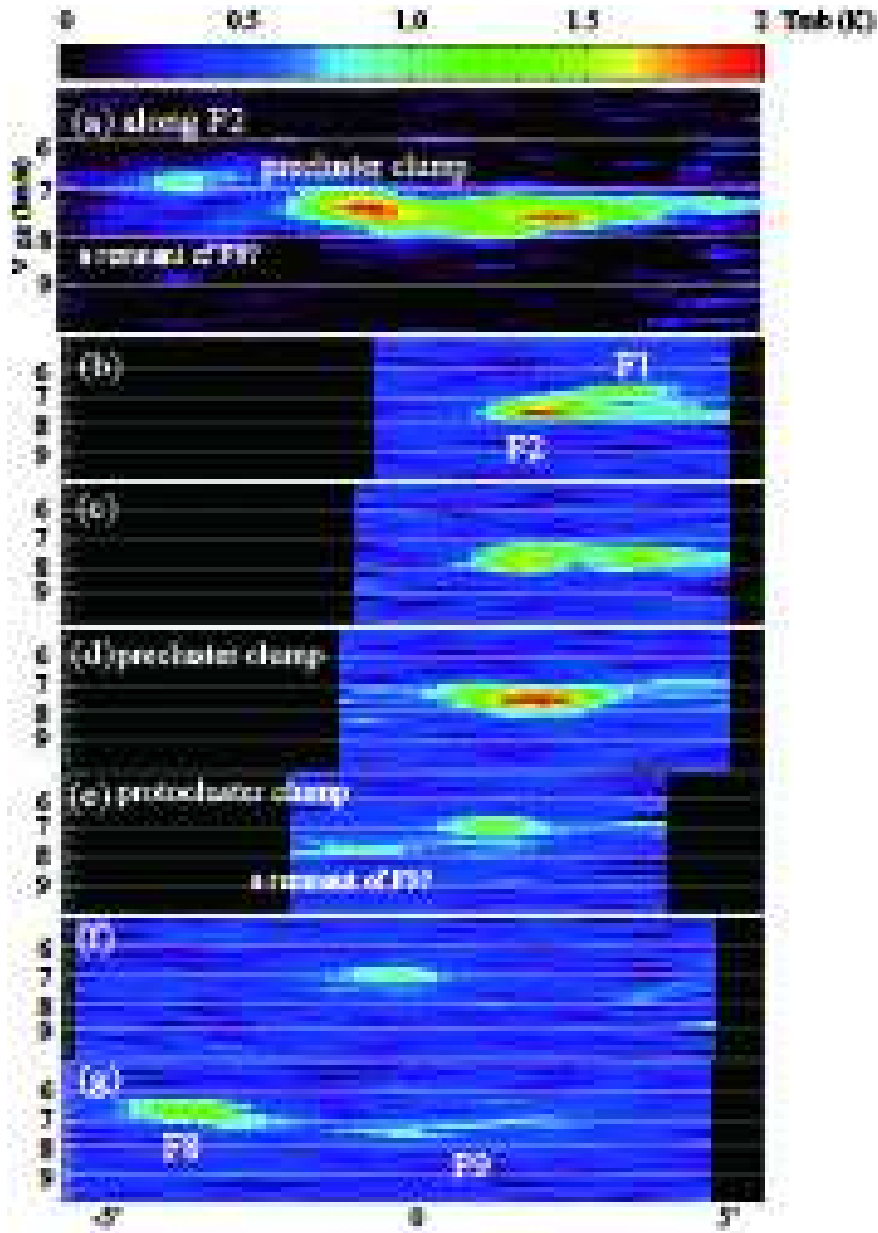}
\caption{Position-velocity diagram of CCS.
The positions of these cuts in the map are shown as white dashed lines in 
}
\label{fig:pvmap}
\end{figure}

Position-velocity diagrams presented in
Figures. \ref{fig:pvmap} (a) though \ref{fig:pvmap} (c) show two components with 
different $V_{\rm LSR}$ in the
northern part. These two appear to converge at the precluster clump
where line width becomes wider (Figure. \ref{fig:pvmap} (c)). 
Such structure is reminiscent of the kinetic structure generated
by filament collision  \citep{duarte11}.
At the protocluster clump, there also appear two components with different
$V_{\rm LSR}$: 6.5 km s$^{-1}$ and 7.5 km s$^{-1}$. The latter is 
weak and vaguely recognized in the position-velocity diagrams. This
component might be a remnant of F9 due to filament collision. 
A large velocity gradient along F8 appears inconsistent with an accretion
flow model \citep{kirk13} because the accretion flow model predicts that 
the velocity gradient near the clump should
be largest because the flow is accelerated by self-gravity and
at both sides of the clump, infall motions should be observed
(Figure 5d of \citet{nakamura91}). Thus, we conclude that our collision
scenario appears more consistent with kinetic structure of Serpens South than
an accretion model.

Recently, the cloud-cloud collision has paid attention as a promising mechanism for
triggering cluster formation 
\citep[e.g.,][]{tasker09,higuchi10,shimoikura13,fukui14,jimenez-serra14}. 
Serpens South is the nearest IRDC
that contains both protocluster and precluster clumps. Therefore, it is an
interesting region to investigate how the cloud-cloud collision
triggers individual star formation by using high-angular resolution 
interferometric observations such as ALMA.

\acknowledgements

This work is supported in part by a Grant-in-Aid for Scientific 
Research of Japan (24244017) and the National Science Foundation under
Grant (NSF PHY11-25915).
We are grateful to the staffs at the Nobeyama Radio Observatory 
for both operating the 45-m and helping us with the data 
reduction. NRO is a branch of the National Astronomical 
Observatory, National Institutes of Natural Sciences, Japan.

\clearpage

\clearpage

\end{document}